\documentclass[sigconf]{acmart}
\usepackage{float}
\newcommand{\myhl}[1]{\textcolor{black}{#1}}

\usepackage{graphicx}
\usepackage{multirow}
\usepackage{footnote}
\usepackage{subfigure}
\usepackage{enumitem}
\usepackage{tikz}
\usetikzlibrary{arrows.meta, positioning, shapes.geometric}

\usepackage[ruled,linesnumbered]{algorithm2e}
\AtBeginDocument{%
  }

\setcopyright{acmlicensed}
\copyrightyear{2018}
\acmYear{2018}
\acmDOI{XXXXXXX.XXXXXXX}

\acmConference[Conference acronym 'XX]{KDD}{June 03--05,
  2018}{Woodstock, NY}
\acmISBN{978-1-4503-XXXX-X/18/06}
\begin{document}
\begin{sloppypar}
\title[Frequency-aware Adaptive Contrastive Learning for Sequential Recommendation]{Frequency-aware Adaptive Contrastive Learning for Sequential Recommendation}

%
\author{Zhikai Wang}
\affiliation{%
 \institution{Fudan University}
 \country{China}
 }
\email{zhikai_wang@fudan.edu.cn}
\author{Weihua Zhang}
\affiliation{%
 \institution{Fudan University}
 \country{China}
 }
\email{zhangweihua@fudan.edu.cn}
%
%
%
%
%
%


\begin{abstract}
Contrastive Learning~(CL) has emerged as an effective paradigm for enhancing Sequential Recommendation~(SR) by generating informative self-supervised signals through data augmentation. However, existing augmentation strategies often disregard the frequency distribution of items and users, leading to over-perturbation of rare but informative interactions, which damages user intent and harms recommendation quality.
In this work, we propose \textbf{FACL}~(Frequency-aware Adaptive Contrastive Learning), a novel yet lightweight framework for robust and intent-preserving sequential recommendation. FACL introduces a micro-level frequency-aware augmentation strategy that adaptively protects low-frequency items and subsequences during perturbation, and a macro-level frequency-aware reweighting scheme that assigns higher importance to sequences with rarer interactions in the contrastive loss. Together, these strategies strike a balance between diversity and fidelity in generated positive views.
We instantiate FACL on top of mainstream SR models and evaluate it on five public benchmarks. Experimental results show that FACL consistently outperforms state-of-the-art methods, improving recommendation performance by up to 3.8\% while maintaining robustness to rare-item sequences.
\end{abstract}


%

\maketitle

\section{Introduction}
\label{sec:introduction}

Sequential Recommendation (SR) aims to predict the next item a user is likely to interact with based on their historical interaction sequence~\cite{GRU4Rec,SASRec,BERT4Rec,FGNN,GAG,PosRec,bao2023tallrec,yu2022graph,time_lstm,din,IMSR,FeSAIL}. SR plays a crucial role in various real-world applications, such as e-commerce, online advertising, and video recommendation, by modeling both users' short-term preferences and long-term evolving interests~\cite{C2Rec,yin2024dataset,cui2024context}.  

In recent years, contrastive learning (CL) has emerged as a powerful paradigm for enhancing representation learning by maximizing the agreement between different views of the same instance while pushing apart different instances~\cite{SimCLR,SimSiam}. Inspired by its success in computer vision and natural language processing, CL has been introduced to sequential recommendation to improve the robustness and generalization of learned sequence representations~\cite{CL4SRec,DuoRec,MCLRec,ContraRec}. Generally, CL-based SR methods can be categorized into two lines: \emph{data augmentation}-based methods~\cite{hrm,CoSeRec} and \emph{model augmentation}-based methods~\cite{RCL,MCLRec}.  

Data augmentation-based methods~\cite{CL4SRec,S3Rec,CoSeRec} generate different views of a sequence by perturbing the input data, such as randomly cropping, masking, or reordering items in the sequence. These augmented views are encouraged to have similar representations. In contrast, model augmentation-based methods~\cite{DuoRec,HPM} achieve this by injecting noise or dropout at the model level without altering the input sequence. Both approaches aim to learn robust sequence representations by exposing the model to diverse contexts.

Though data augmentation-based methods are simple and intuitive, recent studies have shown that model augmentation tends to produce more robust and consistent improvements~\cite{DuoRec,RCL}. A key reason is that data augmentation may inadvertently disrupt the inherent sequential patterns of user behavior, especially when applied indiscriminately. For instance, dropping, reordering, or masking important items in a sequence can distort or even remove crucial intent signals embedded in the user's interaction history.  

To better understand the limitations of data augmentation in SR, we conduct a fine-grained empirical analysis focusing on the frequency of target items in Yelp. Specifically, we find that data augmentation disproportionately harms the representation of \textbf{low-frequency items}. Since these items already have sparse occurrences and fragile co-occurrence patterns, perturbing them—such as dropping them or shuffling their positions—can destroy their already limited context information. In contrast, high-frequency items, which appear in many contexts, are relatively immune to such perturbations.  

\begin{figure}[t]
    \centering
    \includegraphics[width=0.48\linewidth]{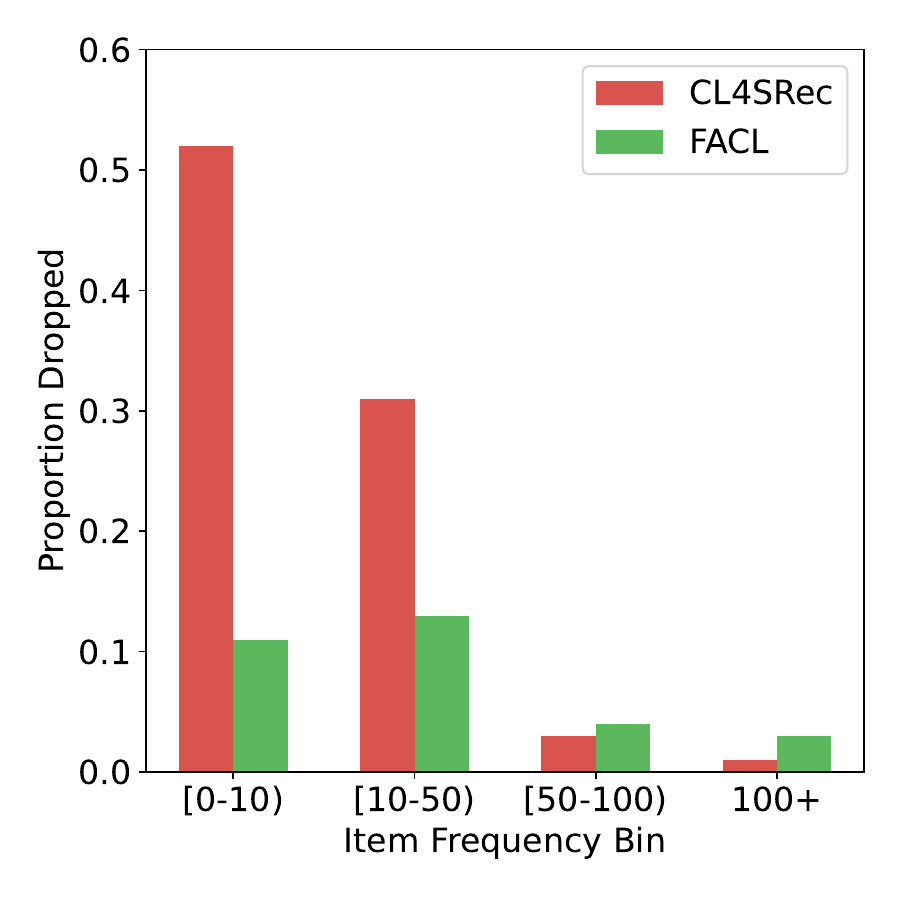}
    \includegraphics[width=0.48\linewidth]{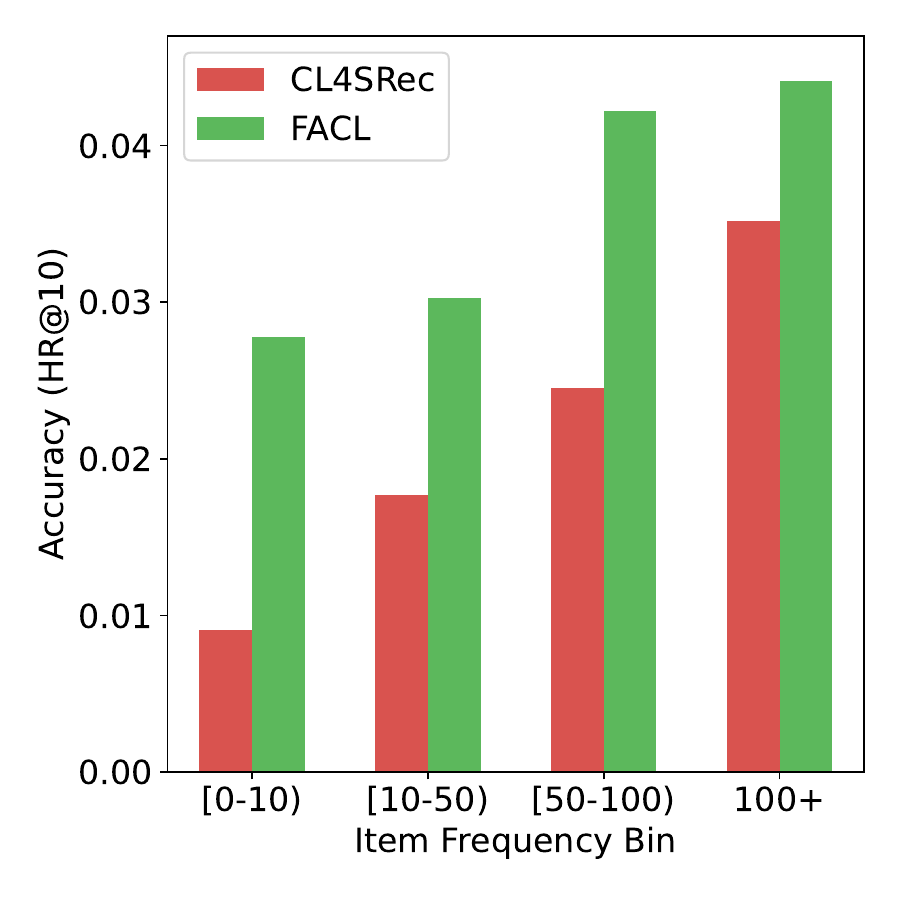}
    \vspace{-0.1in}
    \caption{Left: Proportion of items completely dropped under \texttt{drop-item} perturbation, across item frequency bins (baseline vs. ours). Right: Prediction accuracy of target items at different frequency levels (baseline vs. ours).}
    \vspace{-0.1in}
    \label{fig:drop_ratio}
\end{figure}

Figure~\ref{fig:drop_ratio}(a) illustrates this phenomenon by showing the proportion of items being completely dropped under the common \texttt{drop-item} perturbation across different item frequency bins (e.g., $[0{-}10)$, $[10{-}50)$, etc.). We observe that low-frequency items are much more likely to be dropped compared to high-frequency ones. Furthermore, as shown in Figure~\ref{fig:drop_ratio}(b), the prediction accuracy of low-frequency target items under baseline data-augmentation-based methods~
\cite{CL4SRec} drops significantly, while high-frequency items remain stable. This reveals a critical flaw in existing augmentation strategies: they treat all items equally, ignoring the inherent differences in frequency and informativeness across items.

Motivated by these observations, we propose a novel framework that adaptively controls perturbations for low-frequency items to better preserve their informative patterns in data-augmentation-based CL for SR. Our framework consists of two complementary modules: At the \textbf{micro-level}, we adaptively reduce the degree of perturbations applied to each low-frequency item during data augmentation. This preserves the local sequence structure and reduces the probability of harmful modifications.
At the \textbf{macro-level}, we assign higher training weights to sequences that contain a higher proportion of low-frequency items and, additionally, to those that are relatively shorter in length---since short user sequences indicate sparser user behaviors and are more likely to suffer from a lack of sufficient learning signal.
These two modules work together: the micro-level ensures low-frequency items are not over-perturbed, and the macro-level ensures that sequences characterized by sparsity (either due to rare items or short lengths) contribute more to the learning objective. This dual-level strategy allows the model to preserve and emphasize the subtle but important signals from low-frequency and sparse user behaviors, which are often highly personalized and difficult to capture.

We extensively evaluate our approach on several public sequential recommendation datasets. The experimental results confirm that our method not only alleviates the vulnerability of low-frequency and sparse sequences, but also significantly improves the overall recommendation performance. Notably, our improved data-augmentation-based method even surpasses state-of-the-art model-augmentation-based approaches, highlighting the potential of carefully designed frequency- and sparsity-aware augmentation strategies for robust sequence representation learning.
The main contributions of this work can be summarized as follows:
\begin{itemize}[leftmargin=10pt, rightmargin=0pt]
    \item We conduct an empirical analysis revealing that existing data-augmentation-based contrastive learning methods in sequential recommendation disproportionately harm low-frequency items, which are crucial for personalized recommendations.
    \item We propose an improved contrastive learning framework that adaptively controls perturbations of low-frequency items, significantly reducing their vulnerability in data augmentation.
    \item We introduce a dual-level strategy: (i) at the micro-level, we reduce harmful perturbations to low-frequency items to preserve their local sequence structure; (ii) at the macro-level, we increase the training weights of sequences with low-frequency items or shorter length to amplify their positive contribution.
    \item We demonstrate through comprehensive experiments that our method consistently improves over state-of-the-art data- and model-augmentation-based approaches, particularly benefiting the prediction of low-frequency target items without sacrificing the performance on high-frequency items.
\end{itemize}

\section{Preliminary}
\subsection{Problem Definition}
Sequential Recommendation (SR) aims to suggest the next item that a user is likely to interact with, leveraging their historical interaction data. Assuming that user sets and item sets are $\mathcal{U}$ and $\mathcal{V}$ respectively, user $u \in \mathcal{U}$ has a sequence of interacted items $S_u=\left\{v_{1,u}, \ldots, v_{{\left|S_u\right|},u}\right\}$. $v_{i,u} \in \mathcal{V}\left(1 \leq i \leq\left|S_u\right|\right)$ represents an interacted item at position $i$ of user $u$ within the sequence, where $\left|S_u\right|$ denotes the sequence length. Given the historical interactions $S_u$, the goal of SR is to recommend an item from the set of items $\mathcal{V}$ that the user $u$ may interact with at step $\left|S_u\right|+1$:
\begin{equation}
\arg \max _{v' \in \mathcal{V}} P\left(v_{{\left|S_u\right|+1},u}=v' \mid S_u\right).
\end{equation}

\subsection{Sequential Recommendation Model}
Our method incorporates a backbone SR model comprising three key components: (1) an embedding layer, (2) a representation learning layer, and (3) a next item prediction layer.
\subsubsection{Embedding Layer}
Initially, the entire item set $\mathcal{V}$ is embedded into a shared space, resulting in the creation of the item embedding matrix $\mathbf{M} \in \mathbb{R}^{|\mathcal{V}| \times d}$. Given an input sequence $S_u$, the sequence's embedding $\mathbf{E}_u \in \mathbb{R}^{\left|S_u\right| \times d}$ is initialized, and $\mathbf{E}_u$ is defined as $\mathbf{E}_u=\left\{\mathbf{m}_{1}\oplus\mathbf{p}_1, \mathbf{m}_{2}\oplus\mathbf{p}_2, \ldots, \mathbf{m}_{\left|S_u\right|}\oplus\mathbf{p}_{\left|S_u\right|}\right\}$. Here, $\mathbf{m}_{i} \in \mathbb{R}^d$ represents the embedding of the item at position $i$ in the sequence, $\mathbf{p}_i \in \mathbb{R}^d$ signifies the positional embedding within the sequence, $\oplus$ denotes the element-wise addition, and $n$ denotes the sequence's length.
\subsubsection{Representation Learning Layer}
Given the sequence embedding $\mathbf{E}_u$, a deep neural encoder denoted as $f_\theta(\cdot)$ is utilized to learn the representation of the sequence. The output representation is calculated as:
\begin{equation}
\mathbf{h}_u=f_\theta\left(\mathbf{E}_u\right)\in \mathbb{R}^d.
\end{equation}

Finally, the predicted interaction probability of each item can be calculated as: 
\begin{equation}
  \hat{\mathbf{y}}=\operatorname{softmax}\left(\mathbf{h}_{u} \mathbf{M}^{\top}\right)\in \mathbb{R}^{|\mathcal{V}|}.
\end{equation}
The training loss used for optimizing the sequential recommendation model is as follows:

\begin{equation}
	\mathcal{L}^{rec}=-\sum_{u \in \mathcal{U}}\log \hat{\mathbf{y}}[v_{|S_u|+1,u}],\label{eq:rec}
\end{equation}
where $v_{|S_u|+1,u}$ denotes the ground-truth target item of user $u$.


\section{Methodology}
\label{sec:methodology}

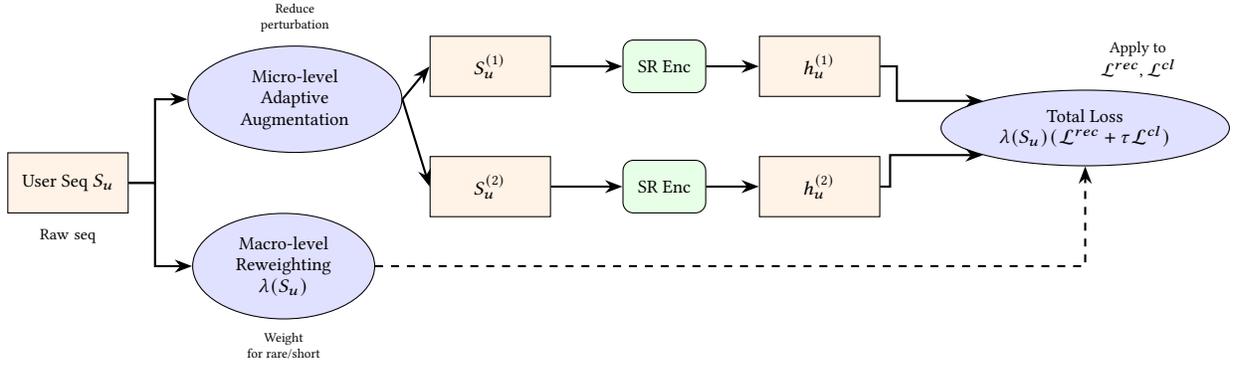
\begin{figure*}[t]
    \centering
    \begin{tikzpicture}[
        font=\footnotesize,
        round/.style={ellipse, draw, minimum height=0.9cm, minimum width=1.4cm, fill=blue!12},
        rect/.style={rectangle, draw, minimum width=1.6cm, minimum height=0.8cm, fill=orange!10},
        process/.style={rectangle, draw, rounded corners, fill=green!10, minimum width=1.1cm, minimum height=0.7cm},
        arrow/.style={thick,->,>=Stealth}
    ]
    \node[rect] (input) {User Seq $S_u$};
    \node[round, above right=0.2cm and 1.2cm of input] (micro) {\parbox{1.8cm}{\centering Micro-level\\Adaptive Augmentation}};
    \node[round, below right=0.2cm and 1.2cm of input] (macro) {\parbox{1.5cm}{\centering Macro-level\\Reweighting\\$\lambda(S_u)$}};
    \node[rect, right=4.0cm of input, yshift=1.55cm] (aug1) {$S_u^{(1)}$};
    \node[rect, right=4.0cm of input, yshift=-0.05cm] (aug2) {$S_u^{(2)}$};
    \node[process, right=0.95cm of aug1] (enc1) {SR Enc};
    \node[process, right=0.95cm of aug2] (enc2) {SR Enc};
    \node[rect, right=0.7cm of enc1] (rep1) {$h_u^{(1)}$};
    \node[rect, right=0.7cm of enc2] (rep2) {$h_u^{(2)}$};
    \node[round, right=0.8cm of rep2, yshift=0.78cm] (loss) {\parbox{2.5cm}{\centering Total Loss\\$\lambda(S_u) (\mathcal{L}^{rec}+\tau \mathcal{L}^{cl})$}};
    \draw[arrow] (input.east) -- ++(0.35,0) |- (micro.west);
    \draw[arrow] (input.east) -- ++(0.35,0) |- (macro.west);
    \draw[arrow] (micro.east) -- (aug1.west);
    \draw[arrow] (micro.east) -- (aug2.west);
    \draw[arrow] (aug1.east) -- (enc1.west);
    \draw[arrow] (aug2.east) -- (enc2.west);
    \draw[arrow] (enc1.east) -- (rep1.west);
    \draw[arrow] (enc2.east) -- (rep2.west);
    \draw[arrow] (rep1.east) -- ++(0.22,0) |- (loss.north west);
    \draw[arrow] (rep2.east) -- ++(0.13,0) |- (loss.south west);
    \draw[arrow, dashed] (macro.east) -- ++(0.7,0) -| (loss.south);
    \node[below=0.1cm of input, font=\scriptsize, align=center, text width=1.7cm] 
      {Raw seq};
    \node[above=0.07cm of micro, font=\tiny, align=center] 
      {Reduce\\perturbation};
    \node[below=0.06cm of macro, font=\tiny, align=center]
      {Weight\\for rare/short};
    \node[font=\scriptsize, above=0.05cm of loss, xshift=0.7cm, align=center] 
      {Apply to\\$\mathcal{L}^{rec},\mathcal{L}^{cl}$};
    \end{tikzpicture}
    \vspace{-0.1in}
    \caption{Overview of the proposed FACL framework.}
    \vspace{-0.1in}
    \label{fig:overview}
\end{figure*}

In this section, we introduce our proposed framework, which improves the robustness of contrastive learning for sequential recommendation by adaptively reducing the perturbation of low-frequency items. As illustrated in Figure~\ref{fig:overview}, our method consists of two key modules: (1) a \emph{micro-level adaptive augmentation module}, which adjusts the probability of perturbing each item or subsequence based on its frequency, effectively preserving the local structure of sequences containing rare items; and (2) a \emph{macro-level reweighting module}, which amplifies the contribution of sequences dominated by low-frequency items during training. Together, these two modules enhance the quality of augmented sequences and alleviate the adverse effects of data augmentation on low-frequency items, enabling the data-augmentation-based contrastive learning approach to outperform even state-of-the-art model-augmentation methods.

\subsection{Micro-level Adaptive Augmentation}

Our empirical analysis (see Figure~\ref{fig:drop_ratio}) reveals that random data augmentation operations tend to disproportionately damage the structure of sequences that contain low-frequency items. For example, such items are more likely to be dropped, substituted, or misplaced, as they lack redundancy in the sequence and in the dataset. To address this problem, we propose an adaptive perturbation strategy that scales down the augmentation probability for low-frequency items, thereby reducing the risk of destroying informative but rare patterns.  

We consider two types of augmentation operations in our design: \emph{subsequence-oriented} and \emph{item-oriented} augmentations.  
Subsequence-oriented augmentations operate on continuous spans of the sequence, e.g., by cropping or reordering subsequences, while item-oriented augmentations operate on individual items, e.g., by dropping, substituting, or inserting items.  
Specifically, we incorporate five augmentation operators, including three random operations (\texttt{crop}, \texttt{drop}, \texttt{reorder}) and two informative operations (\texttt{substitute}, \texttt{insert}) inspired by~\cite{CoSeRec}. The random operators create diverse views by randomly perturbing the sequence, while the informative operators use correlated items to generate more semantically meaningful variations. We next describe the adaptive strategies for these two classes of augmentation.

\paragraph{Item-oriented augmentation.}
Let the global item frequency in the dataset be denoted as $f(v)$ for item $v$, and let $\bar{f}_{S_u}$ be the average frequency of all items in the current sequence $S_u$. We first set a target augmentation ratio $\gamma$, which controls the overall perturbation strength. For each item $v$ in the sequence, we compute a scaling factor based on the ratio between its frequency and the average frequency within $S_u$:
\begin{equation}
    \rho(v) = \gamma \cdot \frac{f(v)}{\bar{f}_{S_u}}
    \label{eq:perturb-prob}
\end{equation}
We then use $\rho(v)$ as the perturbation probability for item $v$, ensuring that the expected augmentation ratio remains at $\gamma$, while protecting low-frequency items (with $f(v) \ll \bar{f}_{S_u}$) from being overly perturbed. To avoid excessively perturbing high-frequency items, we cap \(\rho(v)\) to at most \(2\gamma\).

For a sequence $S_u = [v_1, v_2, \dots, v_n]$, the adaptive per-item perturbation probability of item $v_i$ is denoted by $\rho(v_i)$. The three item-oriented augmentation operators are formally defined as follows.

\textbf{Drop:}
Each item $v_i$ is dropped independently with probability $\rho(v_i)$. The resulting sequence is:
\begin{equation}
S_u^{\text{drop}} =
\big[\, v_i \;\big|\; v_i \in S_u, \; z_i > \rho(v_i) \,\big]
\label{eq:drop}
\end{equation}
where $z_i \sim \text{Uniform}(0,1)$.

\textbf{Substitute:}
Each item $v_i$ is replaced with a correlated item $\tilde{v}_i$ with probability $\rho(v_i)$. The resulting sequence is:
\begin{equation}
S_u^{\text{sub}} =
\big[\, v'_i \;\big|\;
  v'_i = \begin{cases}
    \tilde{v}_i, & z_i \leq \rho(v_i) \\
    v_i, & \text{otherwise}
  \end{cases}, \; z_i \sim \text{Uniform}(0,1) \,\big]
  \label{eq:substitute}
\end{equation}
where $\tilde{v}_i$ is sampled from the set of items most correlated to $v_i$ based on the hybrid correlation score~\cite{CoSeRec}.

\textbf{Insert:}
For each item $v_i$, with probability $\rho(v_i)$ we insert a correlated item $\tilde{v}_i$ immediately after $v_i$. The resulting sequence is:
\begin{equation}
S_u^{\text{ins}} =
\text{Interleave}\bigg( 
  [v_1, \dots, v_n], \;
  \big[\, \mathbb{I}(z_i \leq \rho(v_i)) \cdot \tilde{v}_i \,\big]_{i=1}^n 
\bigg)
\label{eq:insert}
\end{equation}
where $z_i \sim \text{Uniform}(0,1)$ and $\mathbb{I}(\cdot)$ is the indicator function. Here the interleaving inserts $\tilde{v}_i$ after $v_i$ if $\mathbb{I}=1$, otherwise nothing is inserted.

\paragraph{Subsequence-oriented augmentation.}
We sample a continuous subsequence $S'$ of length $c = \lceil \eta \cdot n \rceil$, where $\eta$ is a hyperparameter controlling the proportion of the sequence. The starting position $i$ is uniformly sampled from $[1, n-c+1]$, yielding:
\begin{equation}
S' = [v_i, v_{i+1}, \dots, v_{i+c-1}]
\label{eq:subseq-sample}
\end{equation}

To ensure that the sampled subsequence contains sufficient low-frequency information, we compute its acceptance probability:
\begin{equation}
\alpha(S') = \min\left( \frac{f_{\text{min}}(S')}{\bar{f}},\, 1 \right).
\label{eq:subseq-accept}
\end{equation}
where $f_{\text{min}}(S')$ is the minimum item frequency in $S'$ and $\bar{f}$ is the global average frequency. If $S'$ is rejected (with probability $1-\alpha(S')$), we re-sample until acceptance. 

Once an acceptable $S'$ is obtained, we apply one of the following two operations:
\begin{align}
S_u^{\text{crop}} &= S'
\label{eq:crop}\\
S_u^{\text{reorder}} &=
\big[ v_1, \dots, v_{i-1}, \text{Shuffle}(S'), v_{i+c}, \dots, v_n \big]
\label{eq:reorder}
\end{align}

\subsection{Macro-level Reweighting}
\label{sec:macro-reweight}

While the micro-level adaptive perturbation reduces the risk of corrupting low-frequency items locally, it does not explicitly encourage the model to pay more attention to sequences that are inherently dominated by rare items or to account for the overall sparsity of user behaviors. To address this, we propose a macro-level reweighting mechanism that increases the training weight of sequences with a higher proportion of low-frequency items and further adjusts for the relative sparsity of the user sequence based on its length.

Specifically, for each training sequence $S_u$, we compute both the average item frequency and normalize the sequence length:
\begin{equation}
    \bar{f}(S_u) = \frac{1}{|S_u|} \sum_{v \in S_u} f(v),
    \label{eq:seq-avg-freq}
\end{equation}
where $|S_u|$ denotes the sequence length, and $f(v)$ is the global frequency of item $v$ in the dataset. 

We further denote $\bar{l}$ as the global average sequence length.
Then the reweighting coefficient $\lambda(S_u)$ can be computed as follows:
\begin{equation}
    \lambda(S_u) = \left( \frac{\bar{f}}{\bar{f}(S_u)} \cdot \frac{\bar{l}}{|S_u|} \right)^\beta,
    \label{eq:reweight}
\end{equation}
where $\bar{f}$ is the global average item frequency, and $\beta$ is a tunable hyperparameter controlling the strength of reweighting.

In this way, sequences with a higher proportion of rare (low-frequency) items and/or relatively shorter (thus typically sparser) user histories are assigned larger weights. This adjustment enables the model to better emphasize sequences representing rare behaviors and long-term user interests, which are particularly important in sparse recommendation scenarios. The coefficient $\lambda(S_u)$ is applied to the loss terms associated with $S_u$, amplifying the contribution of such sequences to the overall learning objective.

\subsection{Overall Objective}
\label{sec:objective}

Our framework integrates the standard sequential recommendation loss with the contrastive learning loss computed on augmented sequences, incorporating both micro- and macro-level enhancements.

For each sequence $S_u$, we generate two augmented views $S_u^{(1)}, S_u^{(2)}$ via the adaptive augmentation strategy and denote their representations as $\mathbf{h}_u^{(1)}, \mathbf{h}_u^{(2)}$, and the representation of the target item as $\mathbf{m}_{v^*}$.  

The overall loss is:
\begin{equation}
    \mathcal{L} = \sum_{u \in \mathcal{U}} \lambda(S_u) \Big( \mathcal{L}^{\text{rec}}(S_u) + \tau \cdot \mathcal{L}^{\text{cl}}(S_u^{(1)}, S_u^{(2)}) \Big)
    \label{eq:total-loss}
\end{equation}

The standard recommendation loss is:
\begin{equation}
    \mathcal{L}^{\text{rec}}(S_u) =
    - \log P(v_{|S_u|+1, u} | S_u) =
    - \log \hat{y}_{v^*, u}
    \label{eq:rec-loss}
\end{equation}

For contrastive learning, we use InfoNCE loss:
\begin{equation}
    \mathcal{L}^{\text{cl}}(S_u^{(1)}, S_u^{(2)}) =
    - \log \frac{\exp(\text{sim}(\mathbf{h}_u^{(1)}, \mathbf{h}_u^{(2)}) / \gamma)}{\sum_{u' \in \mathcal{U}} \exp(\text{sim}(\mathbf{h}_u^{(1)}, \mathbf{h}_{u'}^{(2)}) / \gamma)}
    \label{eq:cl-loss}
\end{equation}
where $\text{sim}(\cdot, \cdot)$ is cosine similarity and $\gamma$ is the temperature.

\subsection{Training Algorithm}
\label{sec:train-algo}

We summarize the training procedure of our framework in Algorithm~\ref{alg:adaptive-cl}.

\begin{algorithm}[t]
\caption{Training with Adaptive Contrastive Learning}
\label{alg:adaptive-cl}
\KwIn{Dataset $\mathcal{D}$, augmentation ratio $\gamma$, temperature $\gamma$, 
reweighting exponent $\beta$, total epochs $T$}
\KwOut{Trained model parameters $\theta$}
Initialize model parameters $\theta$ and item frequency table $f(v)$\;
\For{$t=1$ to $T$}{
    \For{each mini-batch $\{S_u\}$}{
        \For{each $S_u$}{
            Compute $\bar{f}(S_u)$ and $\lambda(S_u)$\;
            Generate two augmented views $S_u^{(1)}, S_u^{(2)}$ with adaptive perturbation:\\
            \Indp
            Adjust per-item and per-subsequence probabilities 
            using $f(v)$ and $\bar{f}(S_u)$\;
            Apply one subsequence-oriented and one item-oriented 
            operation per view\;
            \Indm
            Encode $S_u$, $S_u^{(1)}$, $S_u^{(2)}$ to get representations\;
        }
        Compute $\mathcal{L} = \sum_{u} \lambda(S_u) \big( 
        \mathcal{L}^{\text{rec}} + \tau \cdot \mathcal{L}^{\text{cl}} \big)$\;
        Update parameters $\theta$ using gradient descent on $\mathcal{L}$\;
    }
}
\Return{$\theta$}\;
\end{algorithm}

\subsection{Complexity Analysis}

\textbf{Time Complexity.}  
For each sequence of length $n$, computing the per-item frequencies and the sequence-level weight is $\mathcal{O}(n)$.  
Each augmentation operator (drop, substitute, insert, crop, reorder) also runs in $\mathcal{O}(n)$, and the contrastive loss, which compares each positive pair against all negatives in a batch, costs $\mathcal{O}(B^2 d)$ per batch of size $B$ and embedding dimension $d$.  
Therefore, the overall complexity per batch is:
\[
\mathcal{O}(Bn + B^2 d),
\]
which is comparable to standard contrastive learning, as the adaptive components introduce only negligible linear-time overhead.

\textbf{Space Complexity.}  
The additional space comes from storing item frequencies and item correlations, at most $\mathcal{O}(|\mathcal{V}|^2)$, which can be reduced in practice using approximate $k$-nearest neighbors.  
The memory footprint for model parameters and batch computation remains the same as the baseline.

\begin{table}[t]
    \centering
    \caption{Statistics of the datasets after preprocessing.}
    \vspace{-0.1in}
    \setlength{\tabcolsep}{2pt}
    \begin{tabular}{l|rrrrr}
    \toprule
    & \multirow{2}{*}{\# Users} & \multirow{2}{*}{\# Items} & \# Avg.  & \multirow{2}{*}{\# Actions} & \multirow{2}{*}{Sparsity} \\
    &          &          & Length               &            &          \\
    \midrule
    Beauty & 22,363 & 12,101 & 8.9 & 198,502 & $99.93 \%$ \\
    Sports & 35,598 & 18,357 & 8.3 & 296,337 & $99.95 \%$ \\
    ML-1M & 6,041 & 3,417 & 165.5 & 999,611 & $95.16 \%$ \\
    ML-20M & 138,493 & 27,278 & 144.4 & 20,000,263  & $99.47 \%$ \\
    Yelp & 30,499 & 20,068 & 10.4 & 317,182 & $99.95 \%$ \\
    \midrule
    Life & \multirow{2}{*}{2,508,449} & \multirow{2}{*}{276,331} &\multirow{2}{*}{40.8} & \multirow{2}{*}{102,399,201} & \multirow{2}{*}{$99.99 \%$}\\
    Service & & & & & \\
    \bottomrule
    \end{tabular}
    \label{tab:stats}
\vspace{-0.1in}
\end{table}

\section{Experiments}
\label{sec:experiments}
In experiments, we will answer the following research questions:
\begin{itemize}[leftmargin=10pt, rightmargin=0pt]
\item \textbf{RQ1} How does the FACL perform compared with the state-of-the-art methods?
\item \textbf{RQ2} How does each component of the FACL contribute to its effectiveness? 
\item \textbf{RQ3} How do hyperparameters influence the performance?
\item \textbf{RQ4} Where do the improvements of the FACL come from?
\end{itemize}

\subsection{Setup}
\subsubsection{Datasets}
The experiments cover six benchmark datasets, detailed in Table~\ref{tab:stats} after preprocessing:
\begin{itemize}[leftmargin=10pt, rightmargin=0pt]
    \item \textbf{Amazon Beauty} and \textbf{Sports}\footnote{https://jmcauley.ucsd.edu/data/amazon/} utilize the widely-used Amazon dataset with two sub-categories as previous baselines.
\item \textbf{MovieLens-1M/20M} (ML-1M/20M)\footnote{https://grouplens.org/datasets/movielens/1m/} are two versions of a popular movie recommendation dataset with different sizes.
\item \textbf{Yelp}\footnote{https://www.yelp.com/dataset} is widely used for business recommendation. Similar to previous works~\cite{DuoRec,hrm,ema,TAMIC,wang2025triplet}, the interaction records after Jan. 1st, 2019 are used in our experiments.
\end{itemize}

\myhl{We follow the settings of previous works~\cite{DuoRec,MCLRec,HPM,time_lstm,lin2025towards}.
All interactions are treated as implicit feedback, filtering out users or items appearing fewer than five times. The maximum sequence length for the ML-1M/20M datasets is set to 200, whereas for the other four datasets, the maximum sequence length is set to 50.}

We use Top-$K$ Hit Ratio (HR@K) and Top-$K$ Normalized Discounted Cumulative Gain (NDCG@K) across $K$ values of $\{5,10\}$, evaluating rankings across the entire item set for fair comparison, following established methodologies~\cite{MMInfoRec,TiCoSeRec,DHCN,DIM,park2025temporal}.

\begin{table*}[htpb]
    \centering
    \caption{Overall performance. Bold scores indicate the best results (FACL). Underlined scores indicate the second-best (RCL). Improv. shows the relative improvement of FACL over RCL in percentage.}
    \vspace{-0.1in}
\begin{tabular}{c|c|cccccccc|c}
\toprule
Dataset & Metric & SASRec & CL4SRec & CoSeRec & CT4Rec & DuoRec & BSARec & RCL & FACL & Improv.(\%) \\
\midrule

\multirow{4}{*}{Beauty}														
& HR@5    & 0.0365 & 0.0401 & 0.0547 & 0.0575 & 0.0546 & 0.0570 & $\underline{0.0601}$ & $\mathbf{0.0627 \pm 0.0012}$ & 4.33 \\
& HR@10   & 0.0627 & 0.0683 & 0.0772 & 0.0856 & 0.0845 & 0.0850 & $\underline{0.0898}$ & $\mathbf{0.0929 \pm 0.0010}$ & 3.34 \\
& NDCG@5  & 0.0236 & 0.0263 & 0.0353 & 0.0342 & 0.0352 & 0.0362 & $\underline{0.0377}$ & $\mathbf{0.0390 \pm 0.0009}$ & 3.44 \\
& NDCG@10 & 0.0281 & 0.0317 & 0.0431 & 0.0428 & 0.0443 & 0.0448 & $\underline{0.0476}$ & $\mathbf{0.0493 \pm 0.0011}$ & 3.57 \\
\midrule

\multirow{4}{*}{Sports}														
& HR@5    & 0.0218 & 0.0227 & 0.0292 & 0.0311 & 0.0326 & 0.0331 & $\underline{0.0354}$ & $\mathbf{0.0370 \pm 0.0013}$ & 4.51 \\
& HR@10   & 0.0336 & 0.0374 & 0.0458 & 0.0479 & 0.0498 & 0.0505 & $\underline{0.0528}$ & $\mathbf{0.0548 \pm 0.0010}$ & 3.79 \\
& NDCG@5  & 0.0127 & 0.0149 & 0.0199 & 0.0189 & 0.0208 & 0.0212 & $\underline{0.0227}$ & $\mathbf{0.0235 \pm 0.0007}$ & 3.52 \\
& NDCG@10 & 0.0169 & 0.0194 & 0.0244 & 0.0260 & 0.0262 & 0.0270 & $\underline{0.0287}$ & $\mathbf{0.0296 \pm 0.0009}$ & 3.14 \\
\midrule

\multirow{4}{*}{ML-1M}														
& HR@5    & 0.1087 & 0.1341 & 0.1741 & 0.1987 & 0.2038 & 0.2060 & $\underline{0.2113}$ & $\mathbf{0.2202 \pm 0.0015}$ & 4.12 \\
& HR@10   & 0.1904 & 0.2239 & 0.2533 & 0.2904 & 0.2946 & 0.2970 & $\underline{0.3045}$ & $\mathbf{0.3151 \pm 0.0013}$ & 3.55 \\
& NDCG@5  & 0.0638 & 0.0968 & 0.1189 & 0.1346 & 0.1390 & 0.1412 & $\underline{0.1469}$ & $\mathbf{0.1523 \pm 0.0010}$ & 3.66 \\
& NDCG@10 & 0.0910 & 0.1284 & 0.1583 & 0.1634 & 0.1680 & 0.1702 & $\underline{0.1763}$ & $\mathbf{0.1828 \pm 0.0009}$ & 3.65 \\
\midrule

\multirow{4}{*}{ML-20M}														
& HR@5    & 0.1143 & 0.1543 & 0.1902 & 0.2079 & 0.2098 & 0.2140 & $\underline{0.2193}$ & $\mathbf{0.2281 \pm 0.0014}$ & 4.01 \\
& HR@10   & 0.2152 & 0.2323 & 0.2495 & 0.2802 & 0.3001 & 0.3040 & $\underline{0.3113}$ & $\mathbf{0.3226 \pm 0.0017}$ & 3.57 \\
& NDCG@5  & 0.0717 & 0.0995 & 0.1121 & 0.1399 & 0.1428 & 0.1460 & $\underline{0.1503}$ & $\mathbf{0.1564 \pm 0.0011}$ & 4.01 \\
& NDCG@10 & 0.1013 & 0.1302 & 0.1426 & 0.1652 & 0.1735 & 0.1760 & $\underline{0.1805}$ & $\mathbf{0.1869 \pm 0.0013}$ & 3.56 \\
\midrule

\multirow{4}{*}{Yelp}														
& HR@5    & 0.0155 & 0.0233 & 0.0269 & 0.0433 & 0.0429 & 0.0448 & $\underline{0.0481}$ & $\mathbf{0.0501 \pm 0.0005}$ & 4.15 \\
& HR@10   & 0.0268 & 0.0342 & 0.0455 & 0.0617 & 0.0614 & 0.0632 & $\underline{0.0675}$ & $\mathbf{0.0701 \pm 0.0006}$ & 3.85 \\
& NDCG@5  & 0.0103 & 0.0122 & 0.0231 & 0.0307 & 0.0324 & 0.0335 & $\underline{0.0358}$ & $\mathbf{0.0372 \pm 0.0007}$ & 3.91 \\
& NDCG@10 & 0.0133 & 0.0151 & 0.0269 & 0.0356 & 0.0383 & 0.0390 & $\underline{0.0418}$ & $\mathbf{0.0434 \pm 0.0005}$ & 3.82 \\
\midrule

\bottomrule
\end{tabular}
    \label{tab:result}
    \vspace{-0.1in}
\end{table*}

\begin{table}[htpb]
    \centering
    \caption{Performance comparison on FMLP baseline with RCL and FACL. FACL shows consistent improvements over RCL. FACL scores include mean and standard deviation.}
\begin{tabular}{c|c|c c c}
\toprule
Dataset & Metric & FMLP & FMLP+RCL & FMLP+FACL \\
\midrule

\multirow{4}{*}{ML-20M} 
& HR@5   & 0.1384 & 0.2233 & $\mathbf{0.2315 \pm 0.0008}$ \\
& HR@10  & 0.2398 & 0.3144 & $\mathbf{0.3238 \pm 0.0010}$ \\
& NDCG@5 & 0.0925 & 0.1528 & $\mathbf{0.1582 \pm 0.0012}$ \\
& NDCG@10& 0.1283 & 0.1826 & $\mathbf{0.1883 \pm 0.0007}$ \\
\midrule

\multirow{4}{*}{Yelp} 
& HR@5   & 0.0197 & 0.0496 & $\mathbf{0.0513 \pm 0.0006}$ \\
& HR@10  & 0.0321 & 0.0686 & $\mathbf{0.0708 \pm 0.0008}$ \\
& NDCG@5 & 0.0148 & 0.0379 & $\mathbf{0.0393 \pm 0.0009}$ \\
& NDCG@10& 0.0185 & 0.0452 & $\mathbf{0.0466 \pm 0.0005}$ \\

\bottomrule
\end{tabular}
    \label{tab:newbase}
    \vspace{-0.1in}
\end{table}

\subsubsection{Baselines} 
We compare our proposed method with various existing baselines. We consider three categories (I: base model without CL, II: data augmentation-based methods, III: model augmentation-based methods) of comparison methods as follows.
\begin{itemize}[leftmargin=10pt, rightmargin=0pt]
    \item \textbf{SASRec~(I)}~\cite{SASRec} is a single-directional self-attention model. It is a strong baseline in the sequential recommendation.
    \item \textbf{CL4SRec~(II)}~\cite{CL4SRec} uses item cropping, masking, and reordering as augmentations for contrastive learning. It is the first contrastive learning method for sequential recommendation.
    \item \textbf{CoSeRec~(II)}~\cite{CoSeRec} applies contrastive self-supervised learning to sequential recommendation by designing informative augmentation operators that utilize item correlations. CoSeRec aims to generate challenging positive pairs for SSL and shows improved robustness against data sparsity and noise.
    \item \textbf{CT4Rec~(III)}~\cite{CT4Rec} offers a model augmentation-based contrastive learning approach for sequential recommendation, which adds two extra training objectives that ensure consistency in user representations across different dropout masks during training.
    \item \textbf{DuoRec~(III)}~\cite{DuoRec} utilizes both sequences with same target item and model-level augmented sequences as positive samples for contrastive learning.
    \item \textbf{BSARec~(III)}~\cite{BSARec} mitigates the oversmoothing problem of Transformer-based SR models by incorporating Fourier-transformed signals into contrastive representations.
    \item \textbf{RCL~(III)}~\cite{RCL} introduces a relative contrastive learning framework by distinguishing strong and weak samples through a dual-tiered positive selection and a weighted relative loss.
\end{itemize}

\subsubsection{Implementation Details} 
For all methods, including FACL and other baselines, we use SASRec as the backbone model. The embedding size and hidden dimension are both set to 64. The Transformer encoder consists of 2 layers and 2 heads. Dropout rates for the embedding matrix and Transformer module are selected from $\{0.1, 0.2, 0.3, 0.4, 0.5\}$. We use the Adam optimizer~\cite{SASRec} with a learning rate of 0.001 and a batch size of 256.
For hyperparameters specific to FACL, the micro-level augmentation ratio $\gamma$ (see Eq.~\ref{eq:perturb-prob}) and subsequence length ratio $\eta$ is searched from $\{0.1, 0.2, 0.3, 0.4, 0.5\}$, ensuring that data perturbation does not substantially distort user sequences given the sparsity of real-world data. The macro-level reweighting exponent $\beta$ (see Eq.~\ref{eq:reweight}) is tuned over $\{0, 0.05, 0.1, 0.15, 0.2\}$, where $0$ corresponds to no reweighting.
Temperature parameter $\tau$ in the overall loss (see Eq.~\ref{eq:total-loss}) and the InfoNCE contrastive loss (see Eq.~\ref{eq:cl-loss}) is chosen from $\{0.01, 0.05, 0.1, 0.5, 1, 5\}$.
All hyperparameters are selected based on the validation NDCG@10. Each experiment is repeated three times with different random seeds, and the average results are reported.

\subsection{RQ1: Overall Performance}

Table~\ref{tab:result} summarizes the performance of FACL compared with representative baselines across all datasets and metrics. Several key observations can be made.
First, all contrastive learning methods, whether using data augmentation or model augmentation, consistently outperform the vanilla SASRec model. This demonstrates the importance of enhanced supervision for learning better sequential representations compared to training only with next-item prediction.
Comparing the two major types of contrastive learning: \textit{data augmentation}-based methods (e.g., CL4SRec, CoSeRec) and \textit{model augmentation}-based methods (e.g., CT4Rec, DuoRec, BSARec, RCL), we observe that the latter generally achieves stronger results, especially on sparser datasets like Beauty, Sports and Yelp. This is likely because model augmentation better preserves the user’s intent information, whereas data augmentation can inadvertently distort rare or important interactions by randomly cropping, shuffling, or dropping items.
However, our proposed FACL bridges this gap by introducing frequency-aware augmentation and reweighting. As a result, FACL not only outperforms all data augmentation baselines but also surpasses the leading model augmentation methods, across all datasets. The improvement of FACL over RCL is consistent across both sparse datasets (such as Beauty, Sports, and Yelp) and denser datasets (ML-1M and ML-20M), typically ranging from 3\% to 4\% in both HR and NDCG metrics. This demonstrates the effectiveness and robustness of our frequency-aware approach regardless of the dataset characteristics.

As shown in Table~\ref{tab:newbase}, we also examine the robustness of our approach on a non-transformer backbone (FMLP). Both RCL and FACL can significantly boost FMLP’s performance, but FACL maintains a 3–4\% improvement over RCL across all evaluation metrics and datasets.
These results verify that FACL effectively combines the advantages of data and model augmentation approaches, leading to robust gains on both sparse and dense sequential recommendation benchmarks, and across different types of model backbones.

\subsection{RQ2: Ablation Study}

We conduct ablation studies to provide a deeper understanding of each module and augmentation operator in FACL. All experiments are carried out on the ML-20M and Yelp datasets.

\paragraph{A. Effect of Different Augmentation Sets}

\begin{figure}[ht]
    \centering
    \begin{minipage}[b]{0.98\linewidth}
        \centering
        \includegraphics[width=0.95\textwidth]{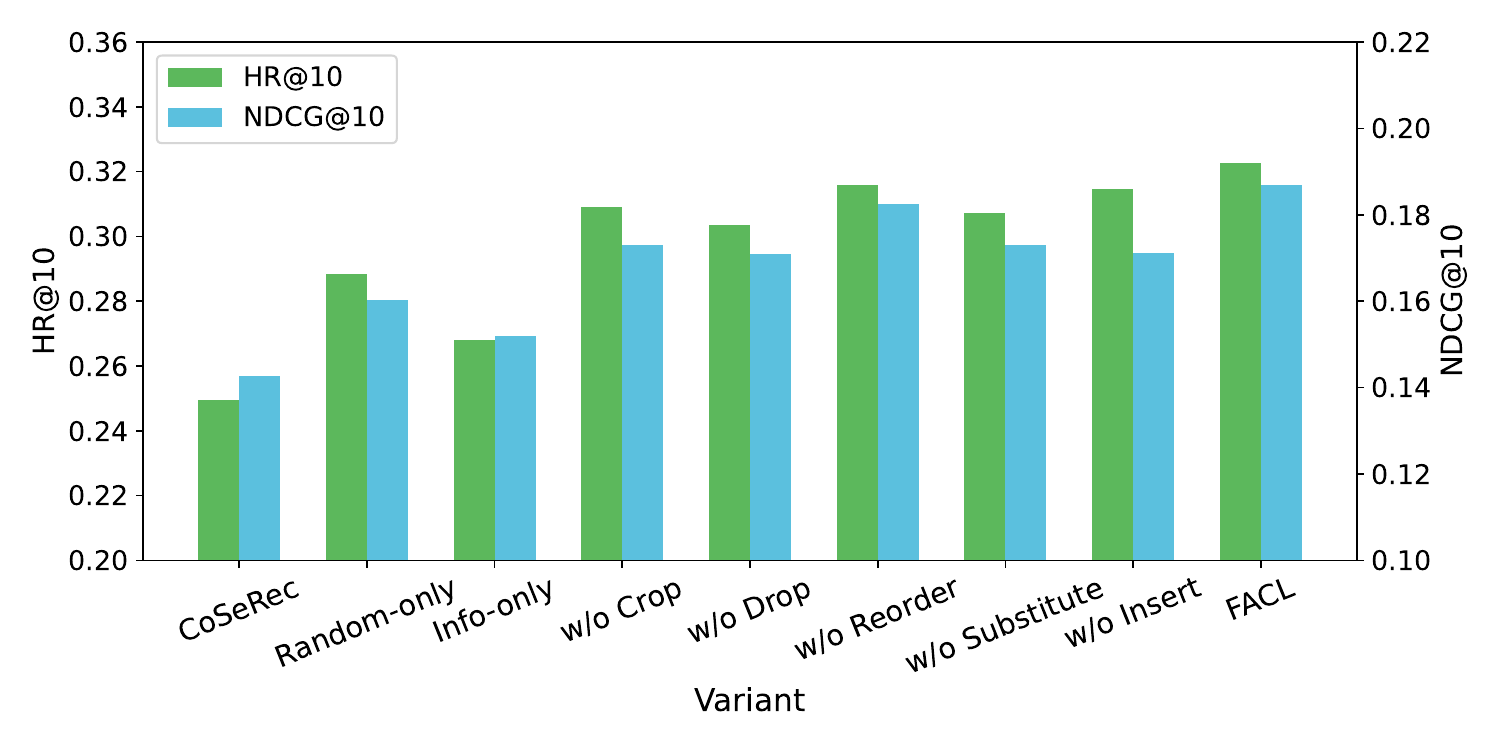}
        \vspace{-10pt}
        \caption*{(a) ML-20M}
    \end{minipage}
    \begin{minipage}[b]{0.98\linewidth}
        \centering
        \includegraphics[width=0.95\textwidth]{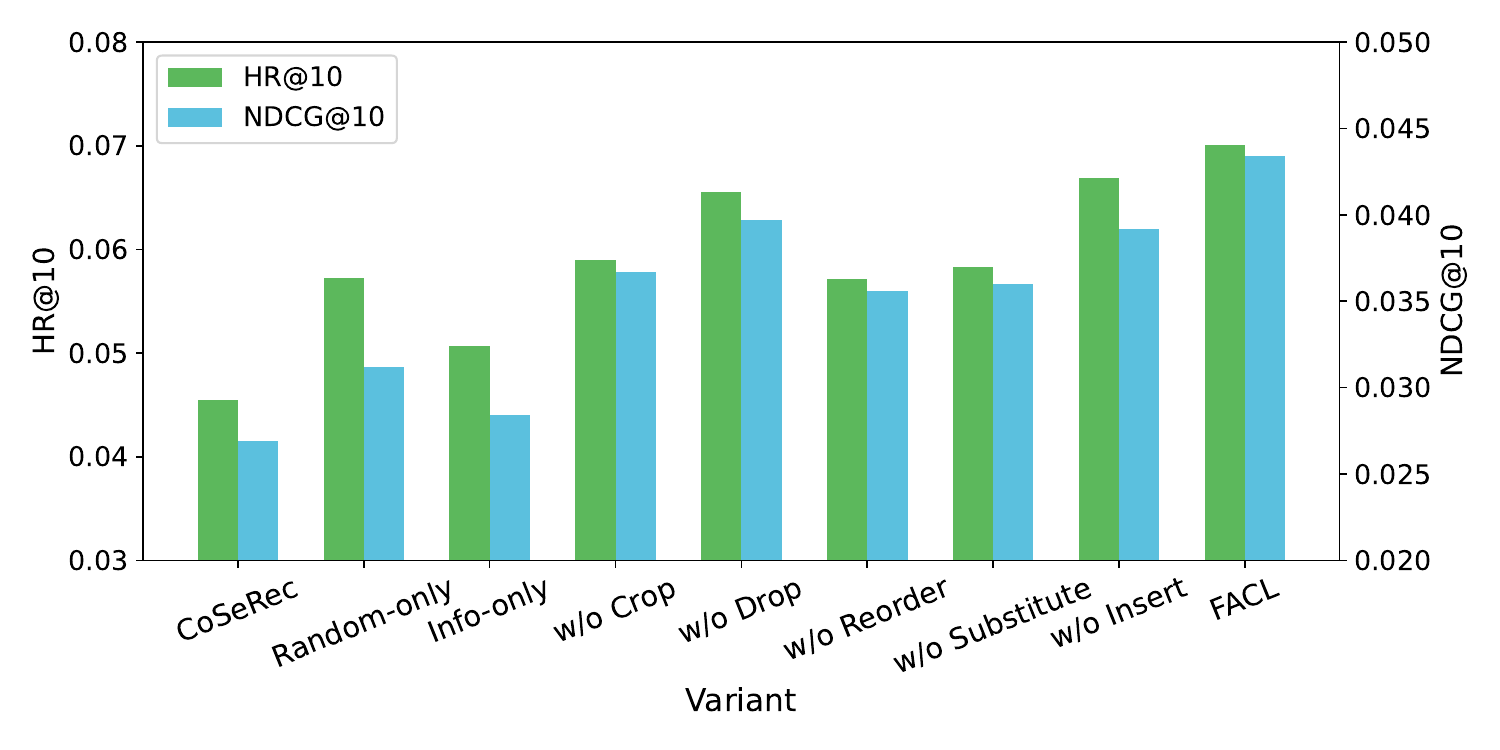}
        \vspace{-10pt}
        \caption*{(b) Yelp}
    \end{minipage}
    \caption{Ablation study of different augmentation sets (CoSeRec variants) on ML-20M and Yelp.}
    \label{fig:ablation-aug}
\end{figure}

We compare FACL with several variants by systematically removing or restricting certain augmentation operators, following the CoSeRec framework. The variants include: (1) CoSeRec (baseline), (2) "Leave-one-out"---FACL with one augmentation operator removed at a time, (3) "Only random aug"---using only random augmentation operators, (4) "Only informative aug"---using only informative operators, and (5) full FACL. Results are shown in Figure~\ref{fig:ablation-aug}.

Key findings include:
\begin{itemize}[leftmargin=10pt, rightmargin=0pt]
    \item All variants outperform CoSeRec, and every Leave-one-out configuration underperforms the full FACL, indicating that each augmentation method contributes meaningfully to overall performance.
    \item Using only random or only informative augmentations yields worse results than using all augmentation types, which means that both random and informative augmentation families are necessary for best results.
    \item The "only random aug" variant outperforms "only informative aug", suggesting adaptive augmentation benefits random perturbations more. This is likely because informative augmentation tends to preserve low-frequency items by using similar items, whereas random augmentation without adaptation risks discarding them entirely.
\end{itemize}

\paragraph{B. Effect of Micro/Macro-level Modules}

To analyze the individual and joint impact of micro/macro-level modules, we conduct ablation studies by disabling or modifying specific weighting strategies in FACL, and compare them against CoSeRec and the full model. Table~\ref{tab:ablation-modules} summarizes the results on ML-20M and Yelp.

We further examine a \textit{len-aware aug.} variant, in which the perturbation probability for each item and each augmented subsequence is additionally scaled by the relative length of that user's sequence. Specifically, the perturbation probability $\rho(v)$ in Eq.~\ref{eq:perturb-prob} and acceptance probability $\alpha(S')$ in Eq.~\ref{eq:subseq-accept} are modified as: 
\begin{align}
    \rho'(v) &= \rho(v) \cdot \frac{|S_u|}{\bar{l}} \label{eq:seq-len-aware-rho} \\
    \alpha'(S') &= \alpha(S') \cdot \frac{|S_u|}{\bar{l}} \label{eq:seq-len-aware-alpha}
\end{align}
where $|S_u|$ is the length of the sequence, and $\bar{l}$ is the average sequence length in the dataset. This adjustment is intended to reflect user sparsity in the data augmentation process.

To evaluate whether LLM-based semantic guidance can serve as an alternative to frequency-based adjustment, we utilize item or user frequency and their textual descriptions as input to a large language model (Gemini 2.5 Pro) to predict the perturbation probability $\rho$, the acceptance probability $\alpha(S')$ and the reweighting coefficient $\lambda(S_u)$, respectively. 

\begin{table}[ht]
    \centering
    \setlength{\tabcolsep}{1.5pt}
    \caption{Ablation results of macro-level modules and sequence-length-aware variants on ML-20M and Yelp. All backbone methods use identical settings as Table~\ref{tab:result}.}
    \label{tab:ablation-modules}
    \begin{tabular}{p{3cm}|cc|cc}
        \toprule
        \multirow{2}{*}{Variant}  & \multicolumn{2}{c|}{ML-20M} & \multicolumn{2}{c}{Yelp} \\
        \cmidrule(lr){2-3} \cmidrule(lr){4-5}
         & HR@10 & NDCG@10 & HR@10 & NDCG@10 \\
        \midrule
        CoSeRec                      & 0.2495   & 0.1426   & 0.0455   & 0.0269   \\
        \midrule
        w/o adaptive aug.            & 0.3142   & 0.1829   & 0.0666   & 0.0411   \\
        w/ len-aware aug.       & 0.3039   & 0.1712   & 0.0621   & 0.0391   \\
        LLM-based aug.       & 0.3077   & 0.1743   & 0.0641   & 0.0396   \\
        \midrule
        w/o reweight                 & 0.3150   & 0.1832   & 0.0674   & 0.0404   \\
        w/o item reweight            & 0.3165   & 0.1841   & 0.0678   & 0.0416   \\
        w/o len-aware reweight     & 0.3169   & 0.1843   & 0.0679   & 0.0417   \\
        LLM-based reweight     & 0.3131   & 0.1816   & 0.0681   & 0.0408   \\
        \midrule
        FACL (full)                  & \textbf{0.3226} & \textbf{0.1869} & \textbf{0.0701} & \textbf{0.0434} \\
        \bottomrule
    \end{tabular}
\end{table}

We observe the following:
\begin{itemize}[leftmargin=10pt, rightmargin=0pt]
    \item Removing either the adaptive augmentation module (\textit{w/o adaptive aug}) or the reweighting module (\textit{w/o reweight}) leads to significant performance degradation, confirming that both are essential for fully exploiting the effectiveness of FACL.
    \item Disabling either the item reweighting or the sequence length reweighting alone also reduces performance. This demonstrates that both user-level sparsity (via sequence length) and item-level rarity must be simultaneously considered when determining the sample weights.
    \item Introducing sequence length awareness into the augmentation probability (\textit{w/ len-aware aug.}, Eq.~\ref{eq:seq-len-aware-rho}-\ref{eq:seq-len-aware-alpha}) results in a consistent drop in performance. This suggests that it is unnecessary to further adjust the augmentation probabilities by sequence length: because the perturbation ratio for each sequence is applied proportionally, shorter sequences are inherently less likely to be over-perturbed.
    \item Additionally, replacing frequency-based adjustment with LLM-predicted reweighting and perturbation probability leads to inferior results. This may be because, as shown in Figure~\ref{fig:drop_ratio}, the effect of data perturbation on accuracy is strongly and negatively correlated with the frequency. LLM-based signals may introduce irrelevant information, leading to less effective adjustment than the direct use of frequency statistics.

\end{itemize}

These results jointly verify the necessity and complementarity of adaptive augmentation and frequency/length-aware reweighting, while also demonstrating that sequence length adjustment in augmentation probabilities is redundant.

\subsection{Parameter Sensitivity (RQ3)}

\begin{figure}[ht]
    \centering
    \includegraphics[width=0.35\linewidth]{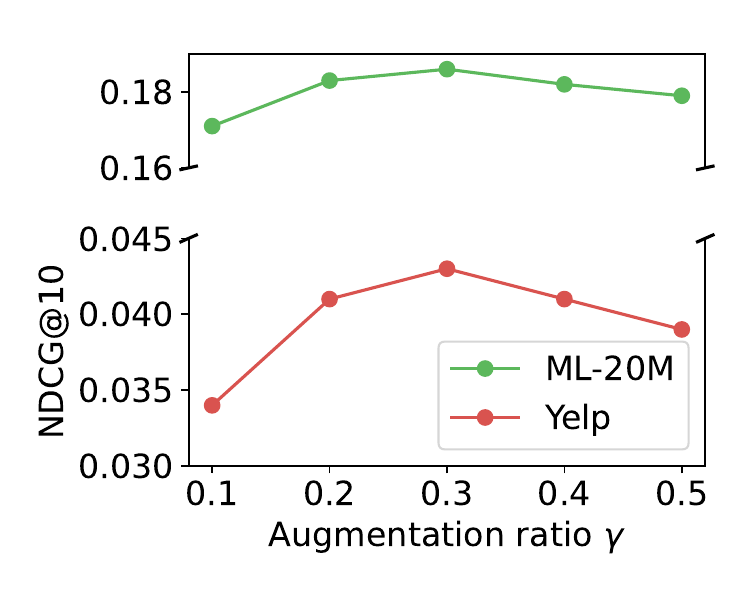}%
    \includegraphics[width=0.35\linewidth]{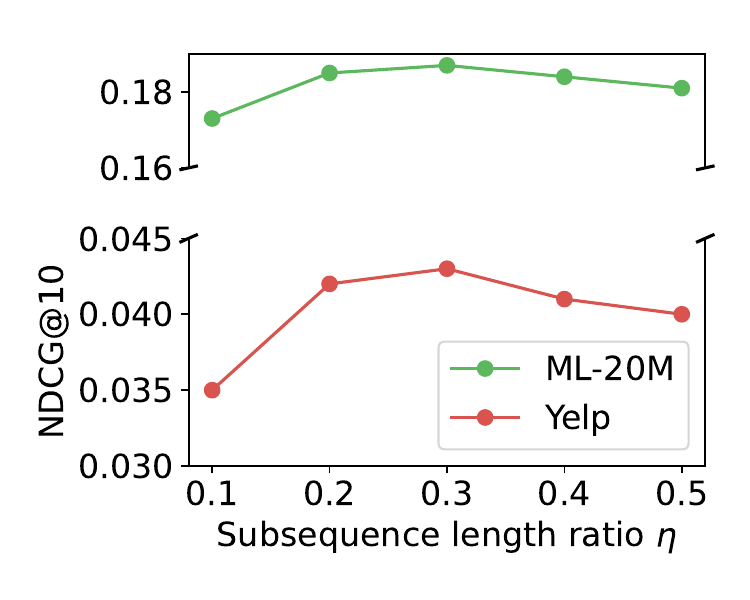}%
    \includegraphics[width=0.35\linewidth]{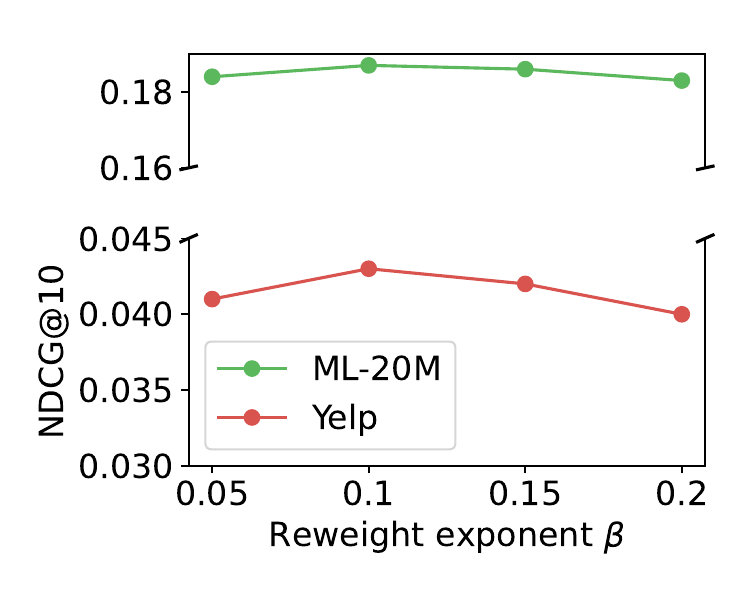}
    \vspace{-0.15in}
    \caption{Parameter sensitivity analysis for (a) $\gamma$, (b) $\eta$, and (c) $\beta$ on ML-20M and Yelp.}
    \label{fig:param-sens}
    \vspace{-0.15in}
\end{figure}

We conduct sensitivity analysis for the main hyperparameters in our framework: the micro-level augmentation ratio $\gamma$, the subsequence length ratio $\eta$, and the macro-level reweighting coefficient $\beta$. Experiments are performed on ML-20M and Yelp, varying one parameter while keeping others at default values; results are visualized in Figure~\ref{fig:param-sens}.

\paragraph{Effect of $\gamma$ (augmentation ratio).}
We vary $\gamma$ from 0.1 to 1.0 to analyze its impact. As shown in Figure~\ref{fig:param-sens}(a), model performance is relatively stable when $\gamma$ is within a moderate range (e.g., $0.2 \sim 0.3$) but drops when set too high or too low. This indicates that an appropriate level of perturbation is important: too little perturbation leads to insufficient self-supervised signal, while too much risks destroying sequence semantics, especially for rare items.

\paragraph{Effect of $\eta$ (subsequence length ratio).}
The hyperparameter $\eta$, as defined in Eq.~\ref{eq:subseq-sample}, controls the ratio of the sampled subsequence length to the full sequence length for augmentation. We find that model performance first improves as $\eta$ increases from small values, reaches an optimum when $\eta$ is moderate (e.g., $0.2 \sim 0.3$), and then saturates or slightly drops as $\eta$ approaches 0.5 (Figure~\ref{fig:param-sens}(b)). Very large $\eta$ yields overly short subsequences, losing sequential semantics; very small $\eta$ results in limited diversity and redundant positive pairs. Thus, a moderate $\eta$ enables effective and meaningful augmentation.

\paragraph{Effect of $\beta$ (reweighting coefficient).}
$\beta$ determines the strength of macro-level (sequence-level) reweighting for rare or sparse sequences. As indicated in Figure~\ref{fig:param-sens}(c), performance peaks at a moderate value of $\beta$ (e.g., $0.05 \sim 0.15$), while too small or too large values lead to loss of benefit or even negative effects. This confirms that controlled amplification of rare sequence contributions is beneficial, but over-weighting may bias the learning process.

Overall, our method exhibits robustness to these hyperparameters in a practical range, with best performance achieved at $\gamma=0.3$, $\eta=0.3$, and $\beta=0.1$ (default settings).

\subsection{RQ4: Item and User Frequency Bias}

To gain deeper insight into the effect of different contrastive learning strategies on items and users of varying frequency, we analyze recommendation accuracy for different frequency groups on ML-20M and Yelp. Specifically, we divide test items and users into several bins (e.g., low, medium, and high frequency) based on their occurrence counts in the training set, and report the hit rate (HR@10) for each group. We compare four representative methods: CL4SRec, CoSeRec, RCL, and our proposed FACL.

\begin{figure}[ht]
    \centering
    \includegraphics[width=0.51\linewidth]{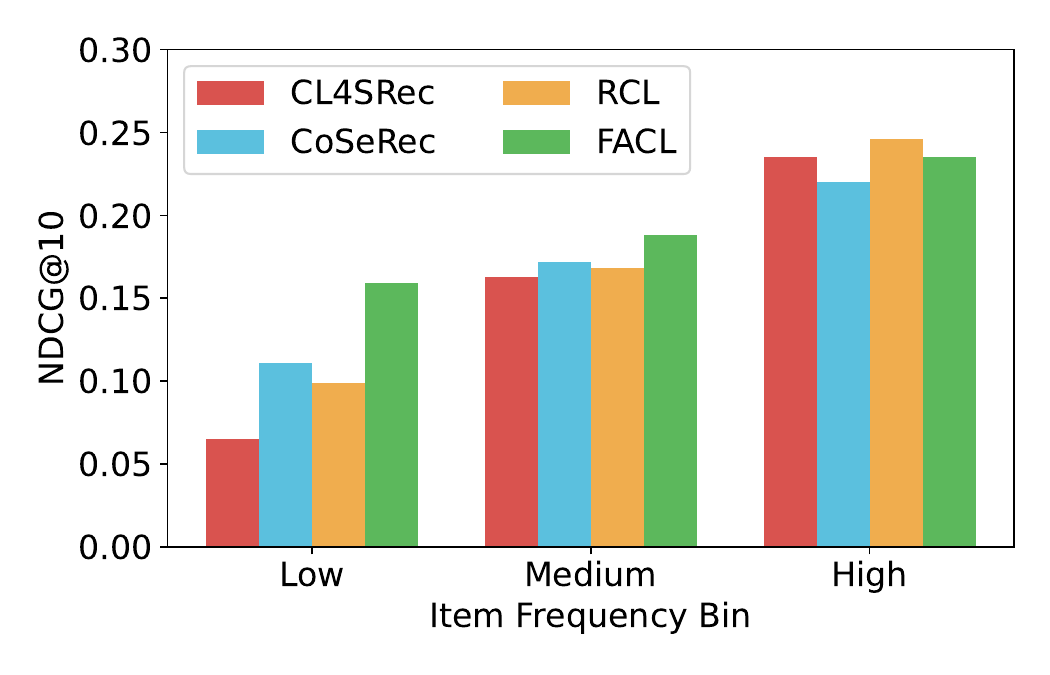}%
    \includegraphics[width=0.51\linewidth]{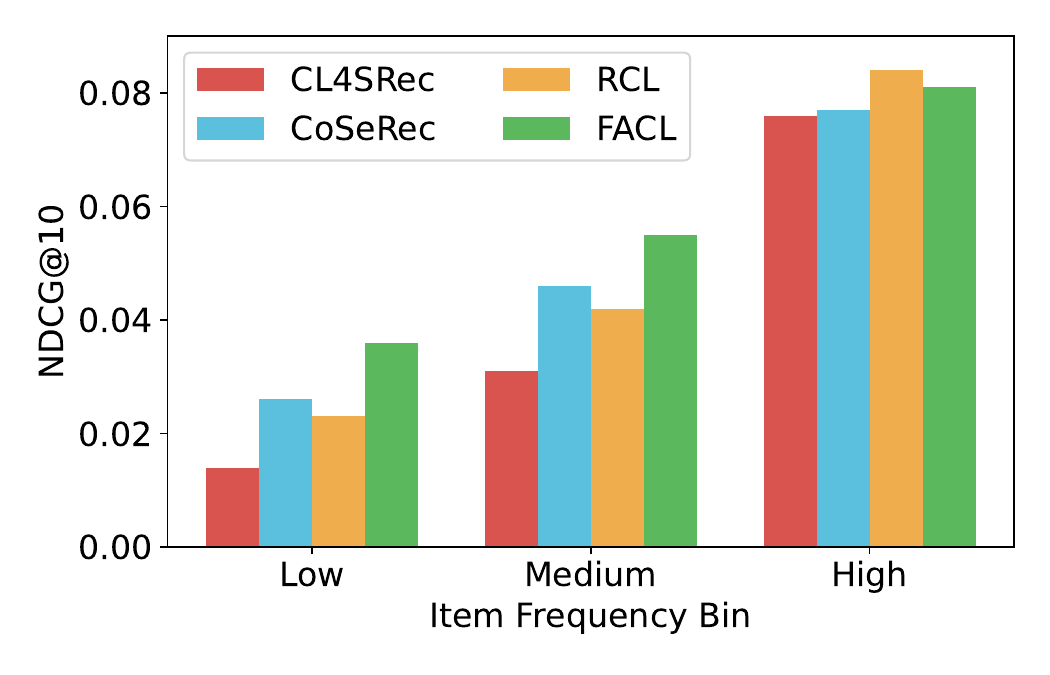}
    \vspace{-0.15in}
    \caption{HR@10 grouped by item frequency (low, medium, high) on ML-20M (left) and Yelp (right).}
    \label{fig:item_freq}
    \vspace{-0.15in}
\end{figure}

Figure~\ref{fig:item_freq} shows the HR@10 across item frequency bins. We observe that for all methods, accuracy on low-frequency (rare) items is lower than that for high-frequency items, due to the inherent data skewness. CL4SRec suffers the most significant accuracy drop on rare items. CoSeRec improves somewhat over CL4SRec, but still exhibits a noticeable drop. RCL alleviates the low-frequency bias, achieving an intermediate performance: its accuracy on rare items is clearly better than CL4SRec, but not as high as FACL. In contrast, FACL achieves the highest accuracy and the smallest gap across item frequencies, indicating its superior capability in mitigating frequency bias and preserving rare item patterns.

\begin{figure}[ht]
    \centering
    \includegraphics[width=0.51\linewidth]{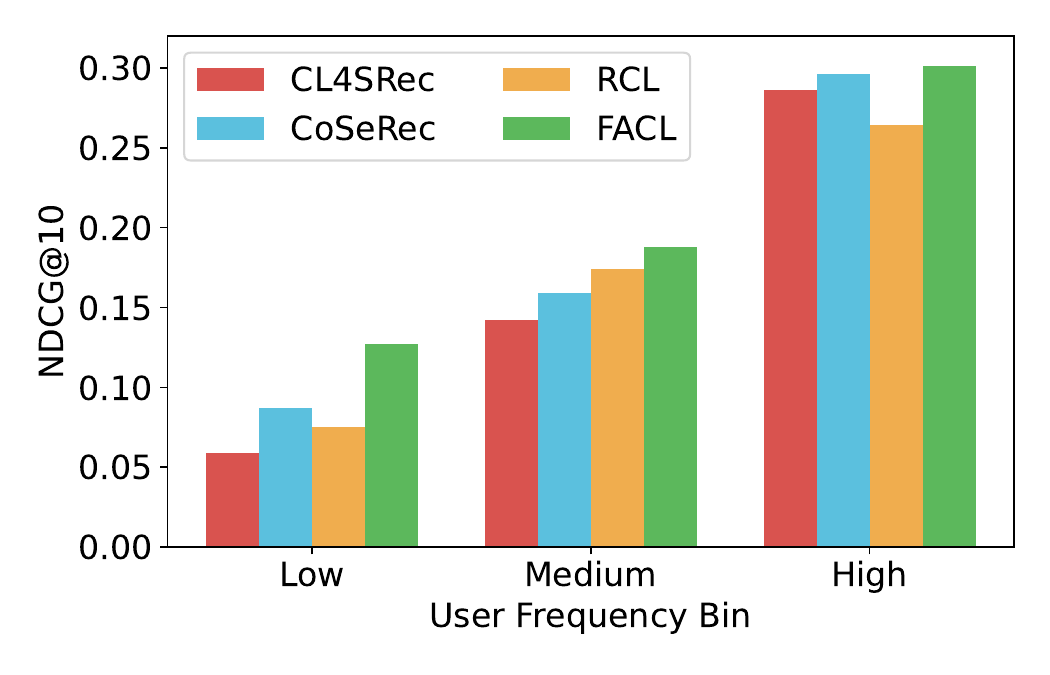}%
    \includegraphics[width=0.51\linewidth]{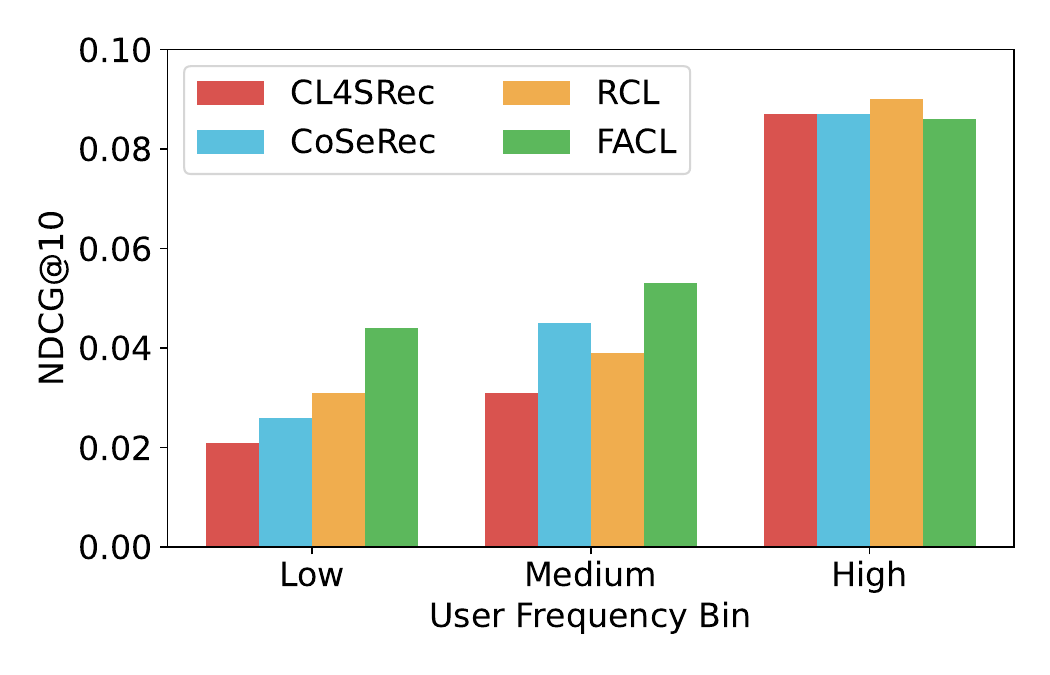}
    \vspace{-0.15in}
    \caption{HR@10 grouped by user frequency (low, medium, high) on ML-20M (left) and Yelp (right).}
    \label{fig:user_freq}
    \vspace{-0.15in}
\end{figure}

Similarly, Figure~\ref{fig:user_freq} presents accuracy over user frequency bins. While accuracy uniformly decreases for lower-frequency (sparser) users, the trends are consistent: CL4SRec has the steepest decline, CoSeRec is slightly better, RCL achieves moderate improvement, and FACL yields the most balanced and robust performance for low-activity users.

Overall, these results demonstrate that FACL not only improves overall accuracy (Table~\ref{tab:result}), but most effectively addresses the major challenge of frequency bias, outperforming both data augmentation and model augmentation-based contrastive methods under sparse conditions.

\section{Conclusion}
\label{sec:conclusion}
In this paper, we revisited the role of data augmentation in contrastive learning for sequential recommendation, revealing its inherent bias against low-frequency items and sparse user behaviors. To address this limitation, we proposed FACL, a frequency-aware adaptive contrastive learning framework that introduces micro-level adaptive perturbation to protect the integrity of rare items, as well as macro-level reweighting to amplify the influence of sparse and rare-interaction sequences during training.
Comprehensive experiments on five public benchmark datasets demonstrated that FACL consistently outperforms state-of-the-art data augmentation and model augmentation-based methods, achieving up to 3.8\% improvement in recommendation accuracy. Moreover, fine-grained analyses confirm that FACL significantly alleviates the performance drop on low-frequency items and users, highlighting its robust intent-preserving ability and its superior applicability to real-world, long-tail recommendation scenarios.

\newpage

\bibliographystyle{ACM-Reference-Format}
\bibliography{referrence}


\end{sloppypar}
\end{document}